# Title: Vibronic resonances facilitate excited state coherence in light harvesting proteins at room temperature

**Authors:** Fabio Novelli[1], Ahsan Nazir[2], Gethin H. Richards[1], Ashkan Roozbeh[1], Krystyna E. Wilk[3], Paul M.G. Curmi[3], and Jeffrey A. Davis[1*]

**Affiliations:**

[1] Centre for Quantum and Optical Science, Swinburne University of Technology, Victoria 3122, Australia.

[2] Photon Science Institute and School of Physics & Astronomy, The University of Manchester, Oxford Road, Manchester, M13 9PL, United Kingdom

[3] School of Physics, The University of New South Wales, Sydney, New South Wales 2052, Australia.

*Correspondence to: JDavis@swin.edu.au

**Abstract**: Until recently it was believed that photosynthesis, a fundamental process for life on earth, could be fully understood with semi-classical models. However, puzzling quantum phenomena have been observed in several photosynthetic pigment-protein complexes, prompting questions regarding the nature and role of these effects. Recent attention has focused on discrete vibrational modes that are resonant or quasi-resonant with excitonic energy splittings and strongly coupled to these excitonic states. Here we unambiguously identify excited state coherent superpositions in photosynthetic light-harvesting complexes using a new experimental approach. Decoherence on the timescale of the excited state lifetime allows low energy (56 cm$^{-1}$) oscillations on the signal intensity to be observed. In conjunction with an appropriate model, these oscillations provide clear and direct experimental evidence that the persistent coherences observed require strong vibronic mixing among excited states.

**Introduction**

Sustaining life on our planet, photosynthesis is one of the most important and studied natural phenomena. In the last few years attention has gathered around pioneering works*(1,2)* that demonstrate signatures of quantum-coherent effects in biological photosynthetic systems*(3,4)*. Subsequently, similar signatures have been found in other light-harvesting complexes*(5-14)* and reaction centres*(15-18)* via multidimensional non-linear spectroscopies.

Notwithstanding the international and multidisciplinary effort, several pivotal open questions remain. Two of the most crucial among these are: what is the physical origin of the oscillations in the non-linear spectral response of these pigment-protein complexes and what is the role, if any, of molecular vibrations*(16-32)*?

In an attempt to clarify these issues and extract precise details of the underlying interactions several experimental and theoretical tools have been deployed. Theories that overcome the limitations of perturbation, Markovian, or adiabatic approaches have been proposed*(24-34)* and experimental approaches that restrict the number of the pathways contributing to the non-linear response have been developed. For example, control of the polarization of each laser pulse*(13,14)* and the order in which they interact*(10)*, or reduction of the bandwidth of the ultra-short laser pulses*(2,6,11-13,18)* have been used to select or exclude specific quantum-mechanical pathways with the aim of identifying the states involved. In each case, however, there remains some ambiguity regarding the nature of the states *(8,9,28,29)* and even whether the coherent superpositions are in the excited or ground state manifold*(10,29)*.

Several groups have proposed models where a vibrational mode is resonant or quasi-resonant with the energy gap between the excitonic states of a dimer and strongly coupled to these states*(25-32)*. The precise implementations vary, and comparisons between the models can be difficult because different bases are used for the observables as well as the eigenstates. However, several important features become apparent from these models. Regardless of whether an explicitly non-adiabatic vibronic basis or an adiabatic exciton-vibrational basis is used, there are two states (the nature of which varies in different models) that are nearly resonant*(25-32)*. The energy difference between these states leads to low energy oscillations that are characteristic of the coupling*(26,27,29,30)*. Another effect of the near resonance is that populations can be coherently transferred between the two states. This opens up several additional signal pathways and makes previously reported means of identifying excited state coherences ambiguous*(29)*. In particular, Tiwari et al.*(29)* showed that it is possible to rationalize most of the long-lived quantum-oscillations detected in 2D experiments on photosynthetic complexes as being due to the coherent superposition of different vibrational levels on the ground state, with their excitation mediated by the coupled vibronic states. Further analysis of the dynamics of these types of model systems has revealed that the strong coupling of the vibrational mode to the electronic states can lead to dephasing times longer than expected for the quantum superpositions of bare excitonic levels*(27-29,31,32)* and can even lead to the generation of coherences from initially excited populations*(27)*.

Recent experimental studies have used coherent multidimensional spectroscopy to explore the role and nature of coherent superpositions in the photosynthetic reaction centre*(16,17)*. In this system both groups were able to identify long–lived coherences that they attributed to coherences of a mixed electronic–vibrational (vibronic) nature. Other work(*18*) has reported coherences in the ground electronic state mediated by the non-adiabatic mixing of the excitonic states and a resonant vibrational mode, as characterised by Tiwari et al.*(29)*. Alternative approaches have studied molecular dimers *(19)* and j-aggregates *(35)* to provide a clearer identification of vibronic coupling and coherences in simpler molecular systems.

Here we demonstrate a novel experimental approach based on two-colour narrowband FWM that allows the relative weight of different quantum pathways to be varied. Specifically, we show that signals arising from excited state coherences dephase on the timescale of the population lifetime and exhibit intensity dependence that matches the saturation behavior of the fluorescent emission, which is significantly different from the intensity dependence of signals arising from ground state vibrational coherences. Utilizing this approach we unambiguously identify excited state coherences in the antenna complex phycocyanin-645 (PC645) from cryptophyte algae at room temperature. Low-energy oscillations on this excited state signal provide direct experimental evidence for the strong mixing between excitonic states and resonant vibrational modes in natural light-harvesting complexes. The frequency of these oscillations matches the energy difference between the resultant vibronic states and our calculations show that such oscillations are present only in the case of vibronic coupling. These observations point to the significance of near resonant vibronic coupling for enabling persistent quantum coherences and inducing non-trivial quantum dynamics in the photosynthetic energy transfer process at physiological temperature(*25-32*).

**Results**
Figure 1 shows the absorption spectrum of the antenna complex PC645 from the cryptophyte algae *Chroomonas* sp. CCMP270*(36-40)*. The two-colour narrowband FWM experiments utilise three pulses with photon energy (wavelength) $E_1$=2.21 eV (560 nm) and $E_2$= $E_3$=2.11 eV (588 nm), selectively exciting coherent superpositions with the first two pulses that are probed by the delayed third pulse (see Methods for details).

**Fluorescence saturation.** Under these excitation conditions, the spontaneous emission of PC645 was also detected, peaking at 1.87 eV*(36-38)* (Fig.1). The fluorescence is primarily a first order process and as such each pulse is expected to contribute independently; hence, we plot the emission as a function of the individual pulse energy (which was kept the same for all pulses) rather than total intensity. Figure 2 shows the fluorescence emission intensity at T = - 400 fs as a function of the excitation intensity of each single pulse (red dots). After an initial linear rise in fluorescence intensity, clear saturation behaviour is observed. A threshold of 6 nJ per pulse, corresponding to 48 ± 4 MW·cm$^{-2}$·nm$^{-1}$ peak power per unit bandwidth per pulse or ~30 μJ·cm$^{-2}$ fluence can be extracted with a simple single-exponential fit.

In general, the transient saturation may be due to depletion of the ground state or filling of the directly excited states, the state/s that radiates, or an intermediate state. In the following paragraph we argue that the saturation of the fluorescence peak shown in Fig.2 is due to the filling of the states directly excited by the laser pulses.

Blocking pulse-1 (leaving just the two lower energy pulses) leads to reduced maximum fluorescence intensity but similar saturation threshold (Fig.2 blue points). If the saturation were in the emitting state the peak fluorescence would remain unchanged because the maximum number of excitons in this fluorescent state doesn't change when one pulse is blocked, allowing us to rule out this possibility. Similarly, since both of the excited excitonic states decay via the same pathways, no change in the peak fluorescence intensity would be expected if the saturation were in an intermediate state. The shared ground state for the excitons suggests that if the saturation were due to ground state depletion (often referred to as a ground state bleach) the saturation threshold would increase as it is due to the total energy, not individual pulse energy, and the peak fluorescence should remain unchanged. In contrast, saturation of the directly excited states leads to reduced fluorescence intensity when pulse-1 is blocked because the total number of photo-excited excitons that can relax to the emitting state is reduced. The saturation threshold, however, remains unchanged because the energy per beam required to saturate the lower energy transition is unchanged. We are therefore able to conclude that the saturation is occurring on the directly excited state.

**Non-linear and intensity-dependent response of PC645.** The detected third-order response of PC645 in our experimental configuration leads to emission with energy centered at $E_4 = -E_1 + E_2 + E_3 \approx 2$ eV. Figure 3(a) shows the signal measured as a function of emission energy and waiting time. The response during pulse overlap (around T=0) is spread over roughly 90 meV and corresponds to the non-resonant third-order response (primarily of the solvent), whereas the signals originating from the dynamic response of the sample are spectrally narrower and extend beyond pulse coincidence. In analysing these plots it is useful to introduce the 'coherence energy', $\Delta$, defined as the difference between the emission energy and the peak energy of pulse-3, and which is equivalent to the energy difference between the two states involved in the coherent superposition.

In Fig.3(b-d) we integrate across this peak from $\Delta = 95$ meV to $\Delta = 115$ meV to show the evolution in T for several different laser intensities. In Fig.3b and Fig.3c we show measurements performed with 37, 60, and 90 MW·cm$^{-2}$·nm$^{-1}$ peak power per unit bandwidth per pulse. Figure 3d highlights the results at the lowest intensity, 11 MW·cm$^{-2}$·nm$^{-1}$.

In a perturbative regime one expects the third-order signal intensity to scale with the third power of the laser pulse intensity. This is indeed the case for the time-zero response, i.e. the non-resonant response induced by the three temporally overlapped driving pulses, as can be seen in the inset of Fig.3b. For the signals that dominate beyond the pulse overlap regime, however, the intensity dependence is not so straightforward.

In Fig.3(c,d) a rescaled y-axis reveals the sample dynamics. At the highest excitation intensities oscillations with a period of $175 \pm 25$ fs are clearly present, corresponding to an energy difference of $24 \pm 3$ meV, which is consistent with a vibrational mode that has previously been identified*(5-8,10-13,39)*. When the excitation density is reduced to 11 MW·cm$^{-2}$·nm$^{-1}$ (1.4 nJ per pulse or 7 μJ·cm$^{-2}$ pulse fluence), the 24 meV oscillations are no longer evident and a slower oscillation with a period of $560 \pm 60$ fs is enhanced (Fig.3d). This slower oscillation is well modelled by a damped cosine function with a single frequency of $7 \pm 1$ meV. This $7 \pm 1$ meV oscillation remains present at greater excitation intensities but it is clear that it scales very differently with intensity compared to the $24 \pm 3$ meV oscillations.

In order to quantify the different dependences on excitation intensity we have performed detailed analyses of these two signal components, as described in the Supplementary Information. The results, reported in Fig. 3e, indicate that the amplitude of the 24 ± 3 meV oscillation scales with the third power of the laser intensity, as expected for a third-order response, whereas the 7 ± 1 meV oscillation displays marked saturation at high photon flux. The saturation threshold of this component is, within our experimental uncertainties, consistent with the value for the excited state saturation observed in the fluorescent emission (Fig.2) and indicative of coherence among excited states, as discussed below. The presence of two signal components that scale differently with intensity suggests that there are two types of pathways contributing to the third-order signal. As described in the following paragraphs, we attribute these two types of pathways to ones evolving in excited state coherences over the waiting time, as shown in Supplementary Fig. 1(d-f), and ones evolving in ground state coherences (Supplementary Fig. 1(b,c)).

**Discussion**

For a given pathway, the saturation threshold (i.e., the pulse energy at which saturation effects become significant) depends on the transition probability of each light-matter interaction and the density of states for each state involved. In a complex system, such as PC645, separating out different pathways is challenging. However, by comparing the saturation observed in the FWM measurements with the saturation of the fluorescence under the same excitation conditions we can gain some understanding. The saturation of the fluorescence involves filling of the excited states, as discussed above. The similar saturation threshold determined for the component of the FWM signal that incorporates the 7 ± 1 meV oscillation strongly suggests that the saturation observed in the FWM measurements is due to pathways evolving in excited state coherent superpositions over the waiting time.

Recent theoretical studies of the intensity dependence of non-linear signals in regimes with relatively high pulse energy have revealed similar effects*(33,34)*. Indeed, the authors of Ref.*(33)* state that "the pulse strength can be viewed as a control parameter, which also permits the discrimination between the contributions from the ground and excited electronic states". This is precisely what we see here, where with high pulse energies the signal is dominated by the ground state vibrational modes, but at lower densities the excited states dominate the signal.

In order to identify the origins of the contributions to the non-linear signal an understanding of the state energies is required. Previous works have identified the presence of two excitonic states separated by 87 ± 3 meV originating from the strongly coupled DBV molecules in the centre of PC645(*8,10-13,36-38,40*) and two vibrational modes with energy 105 ± 10 meV and 24 ± 3 meV, respectively(*5-8,10-13,39*).

The ladder of vibrational modes is shown in the bottom half of Fig.4 together with the bandwidth of Raman transitions accessible with the laser pulse bandwidths used. For the results at high laser intensity it is the excitation of these ground state vibrational coherences that dominates the dynamics. The specific vibrations excited are the vibrational states |0,0,1⟩ and |0,1,1⟩, as indicated in Fig.4. The 24 ± 3 meV oscillations are thus attributed to beating between these two states.

For the dynamics at low excitation intensity we consider the excited states of PC645. The energy level scheme for PC645 shown on the left hand side of Fig.4 is obtained by combining our recent experimental results(13) with other measurements and calculations(10,24,36-38). In Ref.(13) the locations of the states were determined by two-colour narrowband FWM experiments, and their nature was identified based on previously reported energies and the results of polarization-controlled experiments that revealed coherences involving states with non-parallel transition dipoles. For simplicity the states were labelled with adiabatic quantum numbers corresponding to the excitonic state and each of the vibrational modes. However, the energies and particularly the energy separations of states are based on experimental measurements that are typically sensitive to the eigenstates.

We determined the eigenstates of the system taking into account the coupled DBV dimer and the two vibrational modes with energy 24 meV and 105 meV using established parameters as detailed in the Methods. On the right hand side of Fig.4 we show the calculated vibronic states for this system and the primary composition of each. The vibronic coupling leads to states that are separated by ~7 meV and which are responsible for the oscillations observed in the experiment.

Based on this vibronic model, we calculate the dynamics expected for initial conditions matching those excited in the experiments. Specifically, an initial coherent superposition involving states $|\psi_2\rangle$, $|\psi_{10}\rangle$ and $|\psi_{11}\rangle$ matches the state of the system after the first two pulses (see right hand side of Fig.4). The evolution of the two quasi-resonant coherences $|\psi_2\rangle\langle\psi_{10}|$ and $|\psi_2\rangle\langle\psi_{11}|$ is plotted in Fig.3f, and is consistent with the signal measured in the experiment. The behaviour observed here well reproduces the experimental data shown in Fig.3d. Oscillations with a period of approximately 560 fs are observed, arising from the vibronic states $|\psi_{10}\rangle$ and $|\psi_{11}\rangle$ separated by 7 meV, together with a modulation depth approaching one. A signal decay time of approximately 280 fs can be reproduced by incorporating an excitonic decay rate of $1/600$ fs$^{-1}$ and a vibrational relaxation rate of $1/3000$ fs$^{-1}$ (see Methods and Supplementary Information for details). By incorporating vibrational coherences in the ground state as part of the initial conditions and varying their relative weight we are also able to qualitatively reproduce the responses seen at higher intensity in Fig. 3c. These results are shown and discussed in the Supplementary Information.

Further support for the vibronic origin of the oscillations is shown in Supplementary Fig. 2, where similar dynamics and intensity dependence are seen for pulse energies resonant with the equivalent states shifted down by one quanta of the 24 meV vibration. If only one of the pulses is shifted up/down in energy by 24 meV, thereby increasing or decreasing the coherence energy away from 105 meV, these low energy coherent dynamics are not detected. Consistent with these observations, Fig.4 shows that as we move down the calculated ladder of states by 24 meV there continues to be pairs of states separated by ~7 meV that are 105 meV above another state involved in the coherent superposition (specifically, $|\psi_1\rangle$, $|\psi_7\rangle$ and $|\psi_8\rangle$).

Comparisons of our experimental results to purely vibrational or purely excitonic models cannot reproduce the 7 meV oscillations. This is most apparent by simply considering the energy separation between states in the vicinity of our excitation. The DBV dimer that we consider here provides the two highest energy excitonic states, separated by $87 \pm 3$ meV. The calculations also show that the next highest energy state arises from one of the MBV chromophores and is a further 50 meV lower in energy*(37)*. Hence, without coupling to vibrations, these excitonic states cannot lead to the observed 7 meV oscillations. Similarly, if we consider only the vibrational modes in the ground state manifold then there are no states consistently separated by 7 meV, even with combinations of these modes. Furthermore, and as discussed previously, the saturation behavior indicates that the 7 meV oscillations are due to states in the excited manifold.

In conclusion, by experimentally exploring the intensity dependence of the third-order response in the light-harvesting complex PC645 we have been able to isolate room temperature coherences that correspond to excited state pathways and which decay on the timescale of the excited state lifetimes. Where previous work*(16-19)* has identified resonances and attributed different energies to vibronic states, we have directly observed the low-energy oscillations of coherence due to beating between such resonances that can only arise in the picture of vibronic coupling. This provides the clearest evidence yet that strong vibronic coupling is essential to describe the coherent dynamics in photosynthetic complexes. Theoretical models based on this type of vibronic coupling have predicted an enhancement in energy transfer efficiency as a result of this coupling and the appearance of coherence even when only populations are excited. The results presented here and the novel methods used to access them greatly enhance understanding of the complex dynamics in the photosynthetic light-harvesting machinery and enable the exciting possibility of now testing the predicted consequences of coherent vibronic coupling.

## Methods

**Details on the experimental technique.** A Titanium:Sapphire amplified regenerative laser 0.4 W average power and 1kHz repetition frequency is split between two two-stage optical parametric amplifiers. The output of each optical parametric amplifier allows for the generation of short pulses in the visible range via sum-frequency generation with residual Ti:Sa output. One optical parametric amplifier produces pulse-1, the other is split in two to generate pulse-2 and pulse-3. Pulse-1, pulse-2, and pulse-3 are focused into the sample and overlap only at the focus. Each pulse has a bandwidth of about 30 meV (8 nm) and duration of 80 fs. The delay between pulse-1 and pulse-2, $\tau$, was set to zero for all experiments and the 'waiting time' delay, T, was varied by scanning the arrival time of pulse-3. With the signal collected in the phase-matched direction $\mathbf{k}_4=-\mathbf{k}_1+\mathbf{k}_2+\mathbf{k}_3$ and the first two pulses set to have different photon energies and no spectral overlap, only signal pathways involving coherent superpositions during the waiting time are excited and subsequently probed (for details see Supplementary Figure 1). The pulse parameters were verified and confirmation that the signals detected originates from the sample were achieved by running scans with just the buffer solution in the sample holder. These results are shown in Supplementary Fig.7 and indicate that the signals detected beyond pulse overlap do originate from the sample.

In order to allow a direct comparison between our experimental parameters and others (Supplementary Table 1), the intensity of each beam is normalized over the pulse duration and bandwidth, i.e. the unit $MW \cdot cm^{-2} \cdot nm^{-1}$ is used. The spot size measured with a beam profiler was ~160 μm diameter, hence 1 nJ of energy per pulse corresponds to 5 $\mu J \cdot cm^{-2}$ of fluence, 62 $MW \cdot cm^{-2}$ of peak power, or 8 $MW \cdot cm^{-2} \cdot nm^{-1}$ peak power per nanometre bandwidth. The intensity (energy) of each pulse is tuned between 11 $MW \cdot cm^{-2} \cdot nm^{-1}$ (1.4 nJ) and 90 $MW \cdot cm^{-2} \cdot nm^{-1}$ (12 nJ) peak power per nanometre bandwidth.

**Sample preparation.** The antenna complex PC645 is obtained from single-cell cryptophyte algae *Chroomonas* sp. CCMP270 as detailed elsewhere(*11*). Given the water absorption, this organism evolved to collect photons in the region of the solar spectrum roughly between 1.9 eV and 2.2 eV (Fig.1) via eight bilins located in each antenna complex. The central chromophores are two strongly coupled dihydrobiliverdin (DBV) molecules, resulting in a doublet of excitonic states split by ∼ 87 meV and named DBV+ and DBV- with excitation energies above 2 eV(*8,10-13,36-40*). The buffer solution with the light-harvesting complex is placed in a 0.2 mm thick quartz cell to give a solution with optical density ∼ 0.3 in the relevant visible range (Fig.1). As the non-linear response of the antenna does not experience detectable degradation between one measurement and the following, the sample position is fixed throughout each scan. This is consistent with what has been reported previously by others(*37*) and by the minimal temperature change expected as detailed in Supplementary Information.

**Modeling.** We consider an excitonic (DBV) dimer, each site of which is coupled to two oscillator modes. Within the dimer single excitation subspace, we can map this system to a single two-level system coupled to two effective oscillators, with Hamiltonian ($\hbar = 1$)

$$H = \varepsilon_1 |1\rangle\langle 1| + \varepsilon_2 |2\rangle\langle 2| + V(|1\rangle\langle 2| + |2\rangle\langle 1|) + \omega_a a^\dagger a + \omega_b b^\dagger b$$
$$+ \left(\frac{1}{\sqrt{2}}\right)(|1\rangle\langle 1| - |2\rangle\langle 2|)[g_a(a + a^\dagger) + g_b(b + b^\dagger)].$$

Here, $|1\rangle$ ($|2\rangle$) represents a state with excitation on dimer site 1 (2) of energy $\varepsilon_1$ ($\varepsilon_2$), which are coupled to each other with strength $V$. Oscillator modes of frequency $\omega_a$ ($\omega_b$) have creation operators $a^\dagger$ ($b^\dagger$) and are coupled to the dimer with strength $g_a$ ($g_b$). We take experimentally relevant parameter values $\varepsilon_1 = 2.12$ eV, $\varepsilon_2 = 2.11$ eV, $V = 0.0435$ eV, $\omega_a = 0.024$ eV, $\omega_b = 0.105$ eV, and set the system-mode couplings $g_a = \sqrt{0.1}\omega_a$ and $g_b = \sqrt{0.04}\omega_b$ to match the Huang-Rhys factors determined in Ref. [37]. The dynamics of the system-mode density operator $\rho$ is determined by solving the master equation:

$$\dot{\rho} = -i[H,\rho] + \left(\frac{\gamma}{2}\right)[2(\sigma_{-1} + \sigma_{-2})\rho(\sigma_{+1} + \sigma_{+2}) - \{(\sigma_{+1} + \sigma_{+2})(\sigma_{-1} + \sigma_{-2}), \rho\}]$$
$$+ \left(\frac{\kappa}{2}\right)(1 + n_a)[2a\rho a^\dagger - \{a^\dagger a, \rho\}] + \left(\frac{\kappa}{2}\right)n_a[2a^\dagger \rho a - \{aa^\dagger, \rho\}]$$
$$+ \left(\frac{\kappa}{2}\right)(1 + n_b)[2b\rho b^\dagger - \{b^\dagger b, \rho\}] + \left(\frac{\kappa}{2}\right)n_b[2b^\dagger \rho b - \{bb^\dagger, \rho\}].$$

Here, we incorporate spontaneous photon emission from the dimer sites at a rate $\gamma = 1/600$ fs$^{-1}$, where $\sigma_{-i} = |0\rangle\langle i|$ is the lowering operator for site $i = \{1,2\}$, $|0\rangle$ is the electronic ground state, and $\sigma_{+i} = \sigma_{-i}^\dagger$. Vibrational relaxation is also included at a rate $\kappa = 1/3000$ fs$^{-1}$ (taken to be the same for both modes), with $n_j = \left(e^{\omega_j/k_B T} - 1\right)^{-1}$ the thermal occupation number of mode $j = \{a, b\}$ at temperature $T = 300$ K, where $k_B$ is Boltzmann's constant.

For the simulated coherence dynamics shown in Fig. 3f we take an initial state $|\psi\rangle_0 = \frac{1}{\sqrt{3}}(|\psi_2\rangle + |\psi_{10}\rangle + |\psi_{11}\rangle)$, matching the state of the system in the experiments after the first two pulses (see right hand side of Fig.4), and plot the evolution of $|\text{Tr}\{(|\psi_2\rangle\langle\psi_{10}| + |\psi_2\rangle\langle\psi_{11}|)\rho\}|^2$, which is proportional to the measured signal. The electric field of the experimentally-detected signal is proportional to the expectation value of the excited coherent superpositions multiplied by the electric field of the third pulse,. As our detection window includes both the $|\psi_2\rangle\langle\psi_{10}|$ and $|\psi_2\rangle\langle\psi_{11}|$ coherences, the resultant intensity measured is proportional to the quantity plotted in Fig.3f

We note that a thermal state of the 24mev mode has a majority of population in the ground state at 300K. The effect of any partial excitation would simply be to shift all of the states subsequently excited up by one quantum of the 24 meV mode.

Further details on the model are given in the Supplementary Information.

**References:**

[1] Engel, G.S. et al. Evidence for wavelike energy transfer through quantum coherence in photosynthetic systems. Nature 446 782 (2007)

[2] Lee, H., Cheng, Y.C., Fleming, G.R. Coherence Dynamics in Photosynthesis: Protein Protection of Excitonic Coherence. Science 316 1462 (2007)

**Acknowledgments:** The authors gratefully acknowledge funding from the Australian Research Council through Discovery Projects DP130101690 and Future Fellowship FT120100587, Swinburne University of technology visiting researcher scheme and The University of Manchester. FN acknowledges F. Fassioli and D. Fausti for helpful discussions and reading of the manuscript.

**Author contributions:**
JD coordinated the project. JD and GR developed the optical system. KW and PC cultured the cryptophyte algae and isolated the protein. FN performed the experiments with the aid of GR, AR, and JD. FN and JD analysed the data. AN performed the modelling with input from JD. JD, FN, AN and PC wrote the manuscript with help from all authors.

**Figures:**

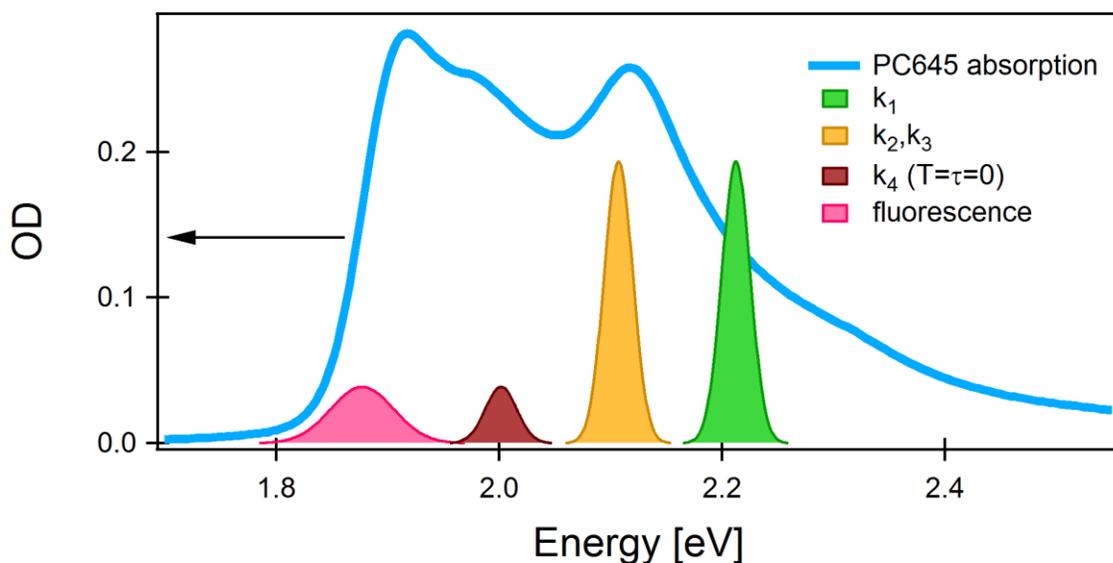

**Fig. 1. Phycocyanin-645 absorption**. Optical density (OD) of 0.2 mm of PC645 with a sketch of the experimental pulse parameters and spontaneous fluorescence.

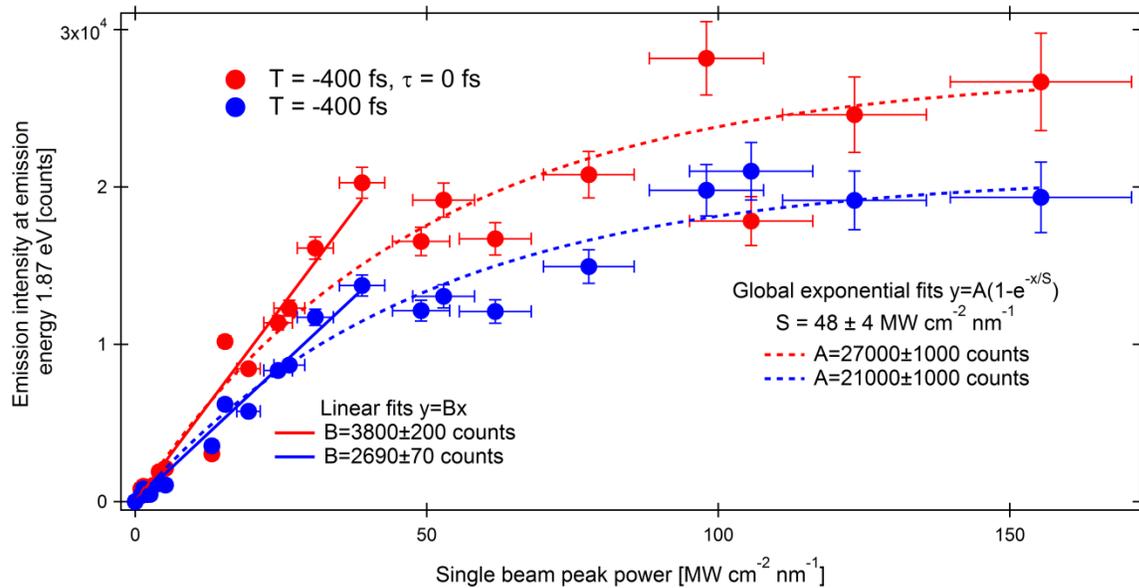

**Fig. 2. PC645 fluorescence saturation.** Emission intensity collected at energy equal to the sample fluorescence peak (1.87 eV) as a function of the intensity of each single pulse. Pulse-2 is delayed by 400 fs from pulse-3. Pulse-1 is blocked in the blue curve, and is at time zero in the red curve ($\tau$=0). The fluorescence saturates versus the intensity of the single pulses either with three (red) or two (blue) beams. Dashed lines are single-exponential global fits that give the critical value of the threshold $S = 48 \pm 4$ MW·cm$^{-2}$·nm$^{-1}$. Solid lines are below-threshold linear fits, confirming that the system response is in the linear regime at low intensities and the ~3:2 ratio of the fluorescent emission amplitudes when three or two beams are used, respectively.

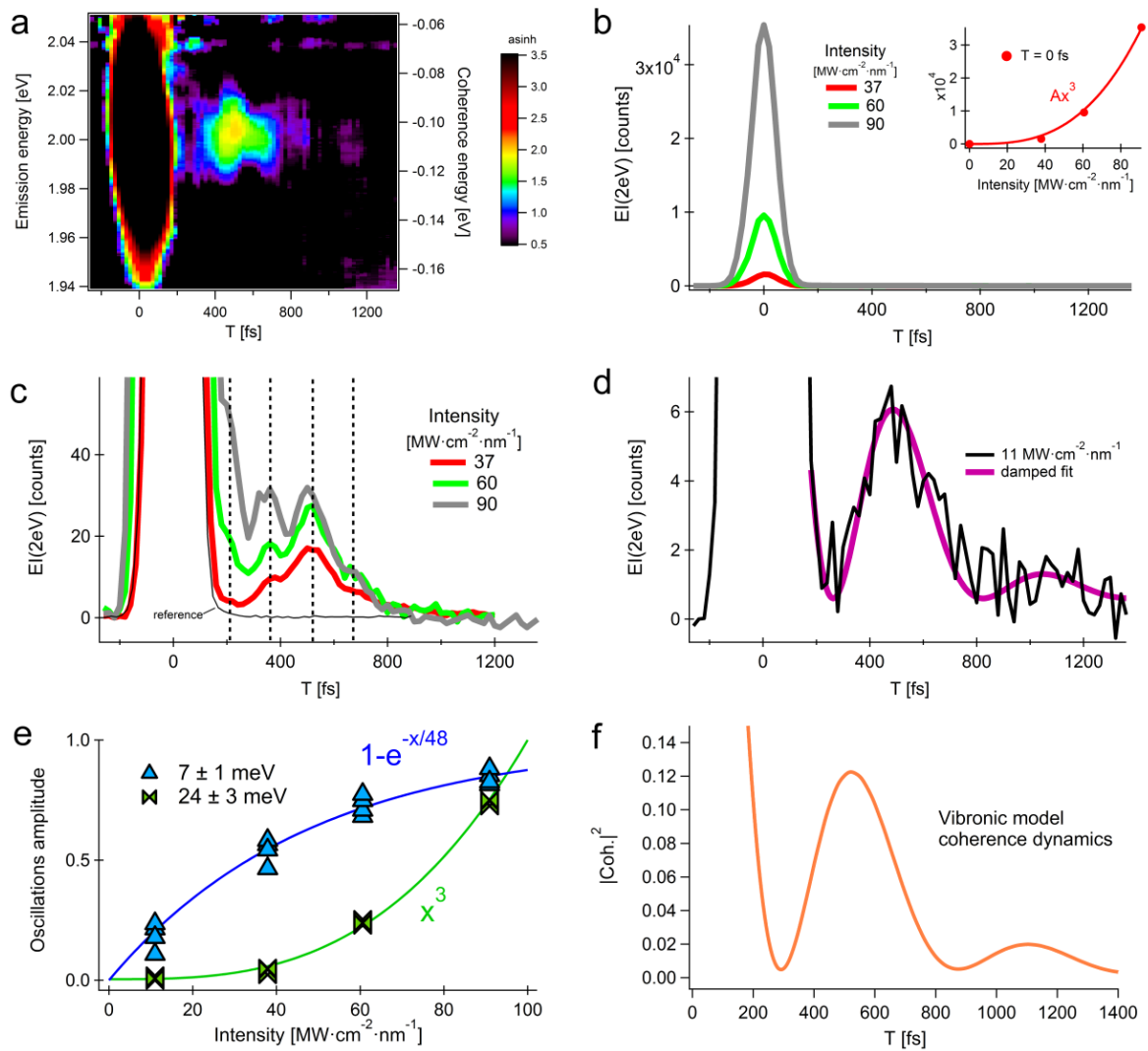

**Fig. 3. Intensity-dependent non-linear response of PC645.** a) Third-order response of light-harvesting phycocyanin-645 at room temperature induced by 80 fs long $E_1=2.21$ eV and $E_2=E_3=2.11$ eV pulses. The emission in the $\mathbf{k}_4$ direction is detected as a function of energy and waiting time. The power of each beam is ~11 MW cm$^{-2}$ nm$^{-1}$. In (b-d), the non-linear signal at 2 eV averaged over a 20 meV interval (± 10 meV) is displayed as a function of the waiting time T. b) Three representative curves for different laser intensities are shown in linear scale. In the inset the third order response at zero time delay is reported. c) Same as b) with an expanded scale. The relative amplitude of the 175 ± 25 fs oscillations (24 ± 3 meV) diminishes with decreasing pulse peak power. The vertical lines are guides for the peak positions. The black curve, obtained on a buffer-only reference at intermediate laser power, does not display any dynamics. d) For low pump intensities only the slower coherence is apparent, well fitted by a simple damped cosine function (purple curve) of the type $Be^{-x/T_B}[1 + cos(\omega_B x + \varphi)] + y_0$. In e) the amplitudes of the two coherent responses are shown, as calculated with different approaches (see Supplementary Information). It is evident that the oscillations with period (frequency) 560 ± 60 fs (7 ± 1 meV) saturate at high laser power with the same threshold of the transient fluorescence (S = 48 ± 4 MW·cm$^{-2}$·nm$^{-1}$). In f) we plot the coherence dynamics $|\text{Tr}\{(|\psi_2\rangle\langle\psi_{10}| + |\psi_2\rangle\langle\psi_{11}|)\rho\}|^2$ (normalized to its initial value) as a function of time for the

initial state $|\psi\rangle_0 = \frac{1}{\sqrt{3}}(|\psi_2\rangle + |\psi_{10}\rangle + |\psi_{11}\rangle)$, calculated from the vibronic model outlined in the Methods.

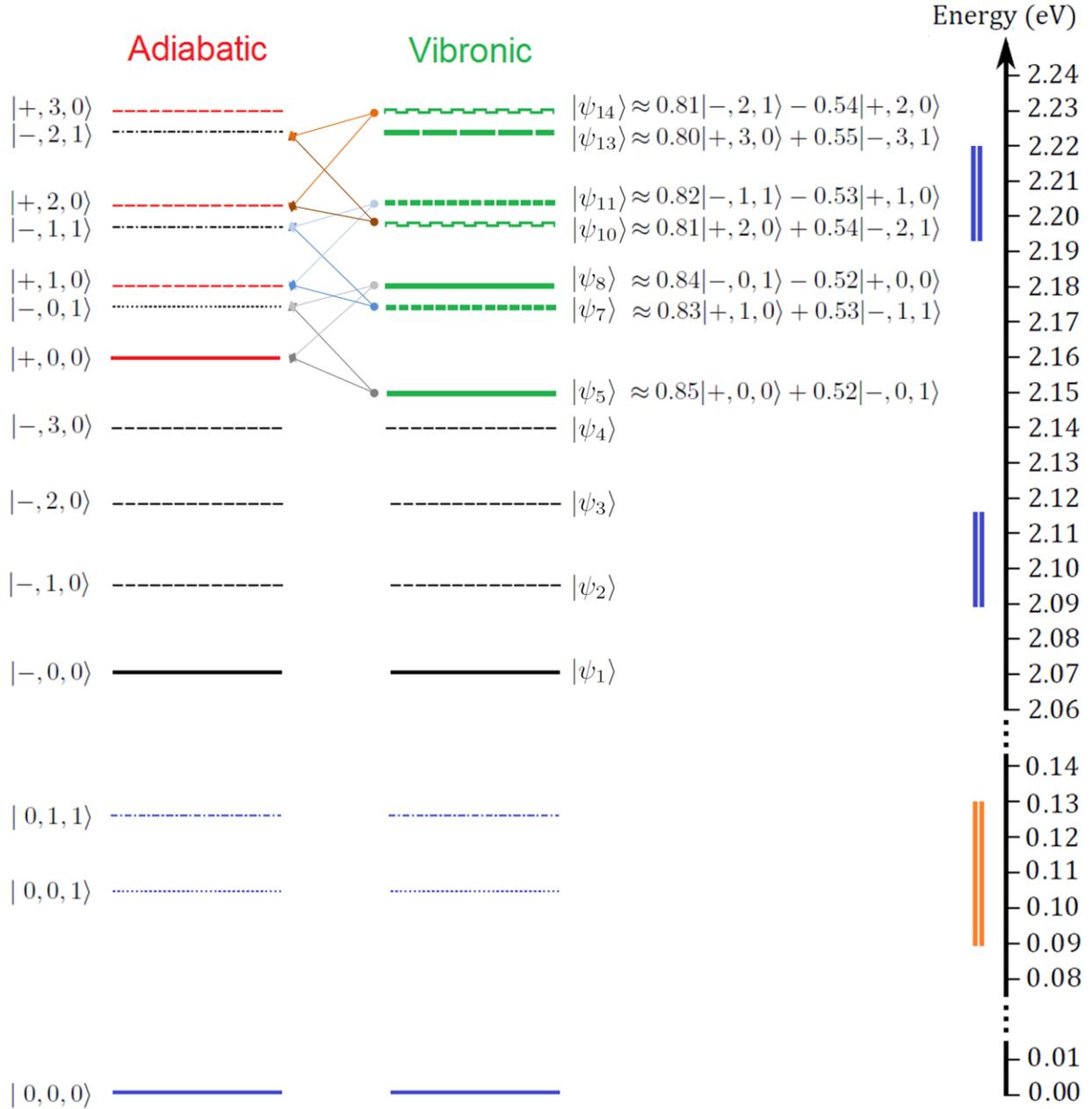

**Fig. 4. Energy level structure of PC645.** The states on the left hand side are obtained by combining previous results*(10-13,24,36-38)*. These states are labeled as |X,Y,Z> with X the excitonic state, Y the ~ 24 meV vibration, Z the ~ 105 meV vibrational mode. The states on the right hand side are the calculated vibronic eigenstates, labeled with dominant contributions from the uncoupled basis states. Given the fact that transition dipole element dominating the transition

from the ground state to a high-overtone of the 24 meV vibration is expected to be vanishingly small, the states $|\psi_6\rangle$, $|\psi_9\rangle$, and $|\psi_{12}\rangle$, have been omitted for clarity. The vertical blue bars represent the ~ 30 meV pulse bandwidth of pulse-1, pulse-2, and pulse-3. The vertical orange bar is $\sqrt{2}$ larger and corresponds to the effective bandwidth for the excitation of vibrations on the excitonic ground state.